# Non-trivial topological spin-polarized surface states induced nonreciprocal transport in the hourglass semimetal

*Zahir Muhammad*[a], *Ghulam Hussain*[b,c], *Zia ur Rehaman*[d], *Rajibul Islam*[a,*], *Igor Antoniazzi*[e], *Muhammad Ismail Khan*[a], *Carmine Autieri*[f], *Maciej R. Molas*[e], *Xiaoguang Li*[b,c], *Xiaoyang Lin*[a,*], *Azizur Rahman*[a,*], *Weisheng Zhao*[a]

[a]National Key Laboratory of Spintronics, Hangzhou International Innovation Institute, Beihang University, Hangzhou 311115, P.R. China

[b]Institute for Advanced Study, Shenzhen University, Shenzhen 518060, China

[c]Institute of Microscale Optoelectronics, Shenzhen University, Shenzhen 518060, China

[d]Nanoscale Synthesis & Research Laboratory, Department of Applied Physics, University of Karachi, Karachi, Pakistan

[e]Faculty of Physics, University of Warsaw, 02-093 Warsaw, Poland

[f]International Research Centre MagTop, Institute of Physics, Polish Academy of Sciences, Aleja Lotników 32/46, PL-02668, Warsaw, Poland

*Corresponding authors: rajibul.physics@gmail.com, XYlin@buaa.edu.cn, azizur@buaa.edu.cn

**Abstract:**

Quantum materials such as topological semimetals have gained consummate interest due to their exotic band crossings and protected surface states. In particular, $Ta_3GeTe_6$ has emerged as a prototypical hourglass Dirac semimetal, as confirmed by both experimental and theoretical analysis. The ARPES confirms the topological hourglass-type band dispersions and Rashba-type band, while Hall measurements reveal a high hole carrier concentration and nonreciprocal transport behavior. These findings are further supported by a phenomenological toy model, which predicts a nonreciprocal resistance, consistent with the observed direction-dependent magnetoresistance. Furthermore, the Raman spectroscopy uncovers two-fold and four-fold phonon symmetries drag of highly anisotropic phonons, which establishes symmetry-dependent resistance. DFT calculations reveal Dirac-like and hourglass Dirac band crossings near the Fermi level for the bulk, with Rashba-type spin splitting induced by the surface inversion-symmetry breaking, leading to spin-polarized surface states. The combined experimental and theoretical results demonstrate that $Ta_3GeTe_6$ hosts a non-trivial electronic topology, with the inversion-symmetry breaking at the surface as the driver of this nonreciprocal transport behavior.

## 1. Introduction

Two-dimensional materials have been under investigation, especially after the realization of a single-layer graphene [1]. The advent of topology in condensed matter opened many new avenues to explore the thrilling physical properties of quantum systems [2, 3]. In this regard, topological semimetals (TSMs) have emerged as an exciting class of quantum materials possessing exotic symmetry-protected Fermions, unmatched and superior electronic properties, unique transport properties, and exceptional carrier mobilities [4-7]. Different classes of TSMs, e.g., Dirac semimetals, nodal-line semimetals, nodal surface semimetals, and Weyl semimetals are gaining popularity since the classification of materials based on symmetry and topology to insulators, metals, or semimetals [8, 9]. Exotic band crossings are a prime feature of topological semimetals, which are ostensibly protected by specific crystalline symmetries, e.g., time reversal symmetry, particle-hole symmetry, etc. Some work has been done to establish a relationship between massless Dirac surface states and their topological protection due to these symmetries. Some of these findings have revealed that in such materials, some anomalous, albeit versatile Fermion states corresponding to crystal lattice symmetries exist [10-11]. These crystal symmetries may result in Dirac Fermion state crossings, hence originating a Dirac point or generating a manifold Dirac nodal point, a Dirac nodal curve or a Dirac nodal loop in k-space. With Dirac Fermions, these materials show exceptional carrier mobility, unique transport behaviour and electrical properties [12-14], making them highly promising candidates for future device applications in areas such as spintronics, high-frequency quantum rectifiers, and directional thermal management.

Recently, the investigations of materials possessing topology-protected electronic properties have become a vital topic of research. Though many TSMs have been proposed theoretically to date, their experimental realization especially for 2D TSMs, is still unusual. In this regard, $Nb_3XTe_6$ (X= Si, Ge) single crystals got enormous attention due to their non-trivial topological properties and hourglass semi-metallic nature of manifold symmetry [15]. They are van der Waals (vdW) materials, which can be exfoliated to achieve a few layers using the mechanical exfoliation technique [16]. The protected symmetry with linear band crossing and Dirac-like dispersions was observed in $M_3XTe_6$ (M = Nb, Ta; X = Si, Ge) compounds [15, 17-19] belonging to centrosymmetric space group No.62 (Pnma). These vdW materials showed the unique topological phase of hosting the weak TI and hourglass TSM characteristics. For this space group, multiple Dirac-like dispersions at the Brillouin zone boundary are robust against SOC, resulting from the interplay of structural anisotropy and nonsymmorphic symmetry [20, 21]. $Nb_3GeTe_6$ is perceived to be a topological semimetal possessing fourfold degenerate nodal-line crossing without spin-orbit coupling (SOC), while an hourglass Dirac loop with the inclusion of SOC [15] in addition to the Dirac-like points. Tuning the SOC interactions in these kinds of compounds can realize various effects. Surface states in topological materials exhibit various SOC physics. The bilinear magnetoelectric resistance of spin-momentum locking and nonreciprocal charge transport is observed in topological materials and other heterostructures attributed to the SOC effect [22-25]. The effect can be observed by applying current and magnetic fields. It shows the resistance can be linearly varied with the applied magnetic field and with the direction of the electric current related to the crystallographic axes of the material. Such a unidirectional magnetoresistance (UMR) was studied in various kinds of non-magnetic and magnetic TIs [22-27]. While recent studies have successfully mapped the bulk electronic structure and nonsymmorphic symmetry-protected bulk nodal lines of $Ta_3GeTe_6$ [28], these investigations typically assume a centrosymmetric bulk framework. However, the critical consequence of surface inversion symmetry breaking, and its associated spin-momentum locking, remains entirely unexplored in this material family. Consequently, the correlation between its topological surface states and nonreciprocal electrical transport has not yet been demonstrated. Owing to the technological importance of TSMs, it is imperative to experimentally understand and systematically explore interesting physical phenomena in such materials.

In this work, we present comprehensive experimental and theoretical investigations of the hourglass Dirac semimetal $Ta_3GeTe_6$, shifting the focus from purely bulk properties to surface-driven quantum transport. High-quality single crystals were synthesized via the chemical vapor transport method and characterized using angle-resolved photoemission spectroscopy (ARPES), Hall transport measurements, and polarized Raman spectroscopy. While our ARPES and Raman data confirm tThe characteristic hourglass-type band dispersions and structural anisotropy of the material, magnetotransport measurements reveal a distinct nonreciprocal transport mechanism. To understand this, a toy model based on Onsager's relation was developed, predicting a nonreciprocal resistance term consistent with the experimentally observed direction-dependent magnetoresistance. Crucially, our first-principles DFT calculations reveal that while $Ta_3GeTe_6$ is a bulk semimetal, spin-polarized Rashba-like surface states with spin–momentum locking emerge natively at the surface due to broken inversion symmetry. Together, these features account for the nonreciprocal transport observed experimentally. Collectively, these results establish $Ta_3GeTe_6$ as a premier platform for exploring surface-mediated nonreciprocal quantum transport, making it a highly potential candidate for future spintronic devices.

## 2. Results and Discussion

High-quality single crystals were synthesized via the chemical vapor transport method (for details, see the method part). The crystal structure was confirmed by using the single-crystal X-ray diffraction method (see Figure S1(b), supplementary information). Whereas the core-level spectroscopy was used to identify the elemental composition of the cleaved $Ta_3GeTe_6$ single crystal surface, with reference to the binding energies (see Figure S1(c), supplementary information). The integrated intensity of each element in the sample can be clearly observed at the corresponding binding position of the present elements. Furthermore, the stoichiometric ratio was measured and confirmed by X-ray spectroscopy (EDS) and the corresponding elemental mapping (Figure S1(d-h), supplementary information). In addition, the crystal structure was measured using scanning transmission electron microscopy (STEM) (see Figure S2a), which forms a zigzag-like structure due to the different oxidation states of Ta atoms with Ge and Te atoms. This structure shows the schematic 1D zigzag structure of the $Ta_3GeTe_6$, having a distorted edge-sharing $TaTe_6$ trigonal prisms interlinked by Ge atoms. The chain propagation vector is a = 6.44758 Å, with out-of-plane vdW stacking axis is b = 13.93736 Å, and inter-chain spacing is c = 11.52903 Å. When projected into quasi-2D TMD-like layers, phase slips (a/2 translational domain boundaries) or grain boundaries between 90°/120°-rotated domain variants interrupt the continuous 1D Ta-Ta motifs, forming localized domain walls (see Figure S2b).

We further conducted ARPES measurements on $Ta_3GeTe_6$ single crystals to explore the electronic properties and band characteristics. Figure 1(a) reveals that the crystal structure of $Ta_3GeTe_6$ belongs to a vdW material family, with Ta and Ge sandwiched between two Te layers. Figure 1(b) represents the 3D first bulk Brillouin zone (BZ) with highlighted high-symmetry points used in the ARPES measurements. In Figure 1(c), the constant energy contour (CEC) at zero binding energy reveals the Fermi surface (FS) with high symmetry directions marked as directed in the BZ. The FS displays a dumbbell-like shape centred at the zone center and expanding to the Y direction, while the open hourglass-like arches (necks at both sides from the zone center from ~ $k_x$ = -0.2 $Å^{-1}$) away from the zone center to the R point of the zone boundary in the first BZ. These features are associated with nodal line bands in these two directions.  The FS was further scrutinized at different binding energies, which shows the asymmetric dispersion around the zone center at R and R' in moment space, revealing the anisotropic feature of the FS (see Figure S3, supplementary information). The CEC plots (Figure S3(a), supplementary information) and the change in position of $k_x$ and $k_y$ in moment space at different binding energies along R and R' points (see Figure S3(b), supplementary information) at the zone boundaries. The results indicate that the $k_x$ and $k_y$ are asymmetric with the anisotropic nature (see Figure S3(c)). These observed momentum-dependent spectral features directly demonstrate electronic anisotropy, consistent with the anisotropic Fermi surface of $Ta_3GeTe_6$.

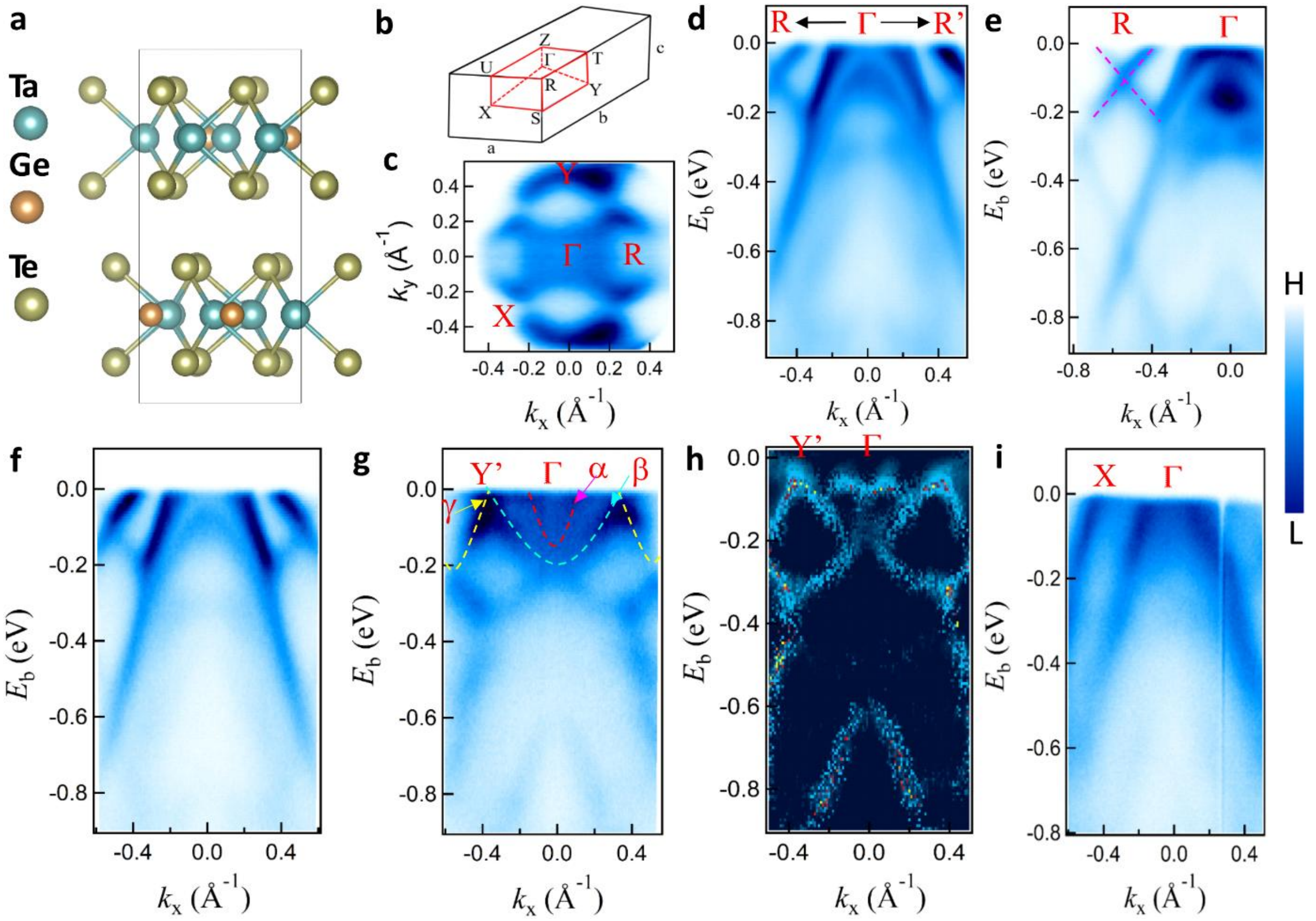

**Figure 1: ARPES measurement for topological electronic properties of $Ta_3GeTe_6$: (**a) Crystal structure of $Ta_3GeTe_6$, (b) Bulk BZ. Fermi surface and band dispersion using ARPES (c) CEC plot at $E_F$ = 0 meV in three different high symmetry directions taken at 20 eV of photon energy, (d & e) cut along R-Γ-R' and R-Γ direction measured at 20 eV with marked nodal lines crossing; (f) cut along R-Γ-R', (g, h) along Y-Γ-Y' with the guide lines representing the shape of the bands and corresponding second derivatives and (i) X-Γ-X' directions, respectively, measured at 24 eV. All the maps and cuts were taken at 8 K.

Figure 1(d) reveals the band dispersion along the R-Γ-R', which illustrates a clear band crossing at the R point making a line crossing band visible at the R point, while the cut is subsequently extended to the R point with a clear X-type band crossing (see Figure 1(e)) at around 100 meV below the Fermi level (FL). Figure 1(e) clearly indicates the sharp nodal lines (seen at ~0.1–0.15 Å$^{-1}$, see Figure S4) band crossing in those specific directions with an intersection at the R and R' point, revealing the nodal lines features of this sample. Figure S4 indicates a band that disperses downward from FL, with the intensity increases in the deeper binding energy, suggesting the band crosses FL somewhere between $k$ =0.0 and $k$ = 0.2 Å$^{-1}$, clear indication of the nodal line band crossing. Figure 1(f) shows the same cut, which was measured at 24 eV, to show both R and R' points for a more visible band crossing. At the same time, the cut along the high symmetry Y'-Γ-Y direction was measured, which shows the different band features (see

Figure 1(g)). In Figure 1(g), the $\gamma$ band reveals the Dirac-type band at the FL along Y and Y' points, which can also be seen in the calculated band structure. The two parabolic bands illustrate the split bands of Rashbha types represented as $\alpha$ and $\beta$. The dispersion of these two bands demonstrates that the bands along the Y-Γ-Y' point possess the dispersive electron-like bands crossing the Fermi level, while at lower binding energy, a hole-like parabolic band touches the $\beta$ band at ~200 meV. The corresponding second derivative (see Figure 1(h)) clearly shows such band features. On the other hand, along X and X' it shows the linear band dispersion, which can cross the FL, revealing (see Figure 1(h)) a clear nodal-line feature similar to $Ta_3SiTe_6$ [19]. These results reveal the linear band crossing along the specified symmetry direction, which is an indication of not only the topological semimetal features but also reveals, these Rashba-like bands, which are spin-polarized, confirmed by the spin-polarized surface calculations as well (see Figure 4(g)). Furthermore, the band dispersion was measured at different photon energies, with varying intensities, affirming the 3D dispersion of the bulk states with $k_z$ dependency in $Ta_3GeTe_6$, as confirmed by the constant energy contour (CEC) map of $k_z$ vs $k_x$ plots (see Figure S5). At different binding energies, the varying intensities of the CEC map along the $k_z$ direction confirm the topological feature of this sample. Whereas the nodal line and Dirac-like bands also show the distinguished band curvature, indicating the anisotropic electronic properties of this material, as we elucidated from the Fermi surface plot as well.

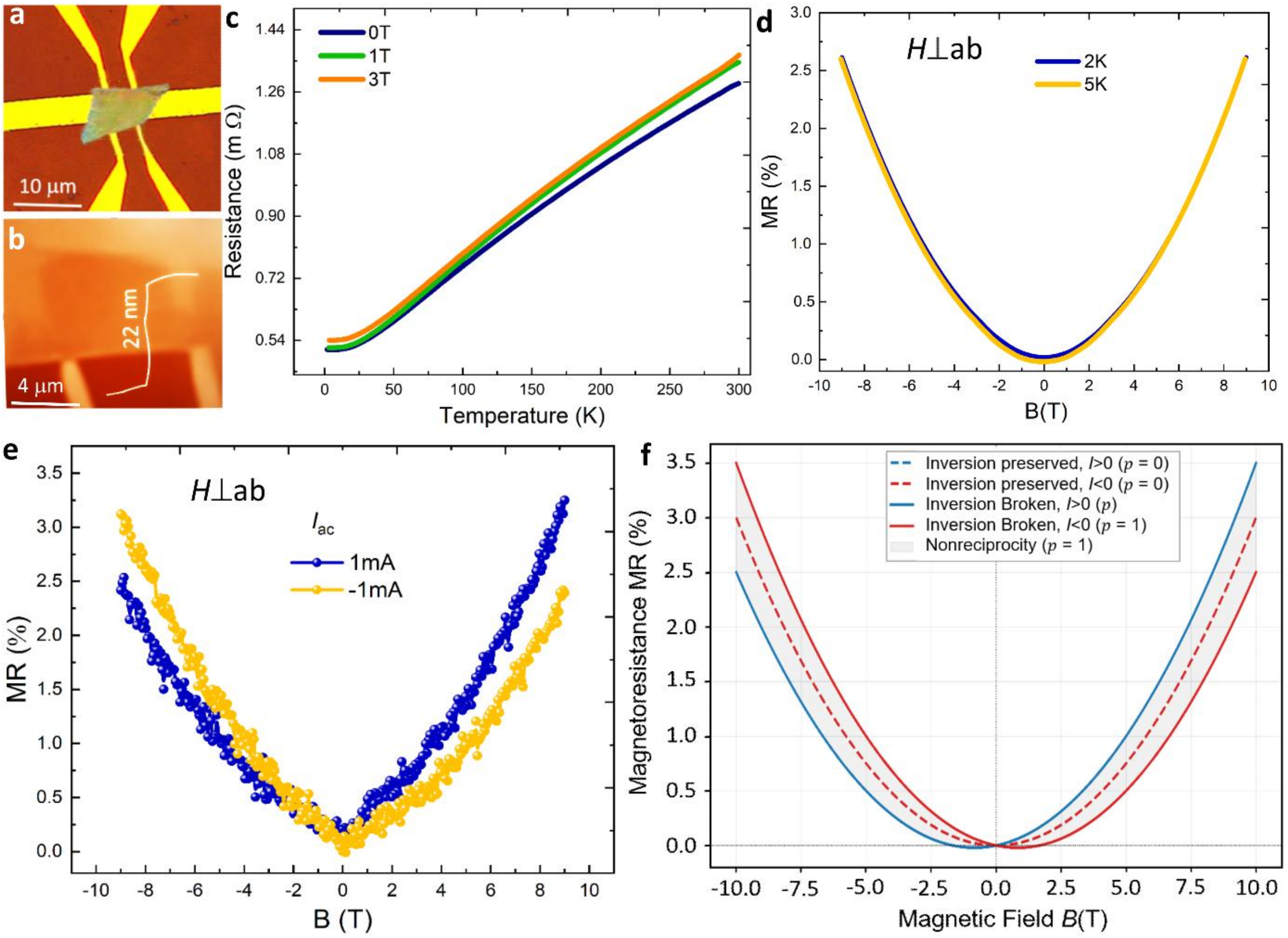

**Figure 2. Transport properties of $Ta_3GeTe_6$ device:** (a, b) Optical image of an exfoliated thin flake of $Ta_3GeTe_6$ on a Hall bar device and the AFM image to confirm the thickness of the $Ta_3GeTe_6$ flake. (c) Measured resistivity of the device as a function of temperature at different magnetic fields. The external magnetic fields were applied perpendicular to the ab-plane (H ⊥ ab, parallel to the c-axis). (d) The magnetoresistance as a function of the magnetic field applied parallel to the c-axis was measured at 2 K and 5 K under an applied DC current ($I_{dc}$), demonstrating symmetric reciprocal transport. (e) Magnetoresistance measured at 2 K under opposite current

directions ($I_{ac} = \pm 1$ mA), showing distinct nonreciprocal splitting. The signal separation between positive and negative current polarities substantially exceeds the experimental noise floor across the entire field range, confirming the intrinsic nonreciprocal response. (f) The magnetoresistance as a function of magnetic field is shown for *I*(+Ve) in blue and *I*(-Ve) in red. The dashed curves correspond to the case with inversion symmetry, while the solid curves represent the scenario with broken inversion symmetry.

The Hall bar device was fabricated (see methods part for details) on the thin flake of $Ta_3GeTe_6$ with a thickness of around ~22 nm, as shown in Figure 2(a) & (b). Figure 2(c) represents the resistivity plot of the $Ta_3GeTe_6$ device as a function of temperature measured at different magnetic fields (***B***) (0T, 1T, 3T). It can be seen that the sample shows metallic behaviour under all tested temperatures and ***B***, with a slight change in resistivity under the applied ***B***. The room temperature resistivity of $\rho_{300K}$ was recorded to be ~1.27 mΩ, while the value of resistivity at the lowest recorded temperature, i.e., at 2 K ($\rho_{2K}$) was found to be 0.45 mΩ. The residual resistivity ($\rho_{300K}/\rho_{2K}$) value is 2.903, which manifests the strong conductivity of the grown crystals. However, the resistance increases with the applied ***B***, having no upturn can be observed in resistance, indicating the typical strong metallic nature of $Ta_3GeTe_6$. Figure 2(d) and Figure S6(a) revealed that the longitudinal magnetoresistance (MR) at different temperatures was measured under a DC applied current ($I_{dc}$) and an applied ***B*** perpendicularly to the *ab* plane or parallel to the *c* plane. It can be noticed that MR is positive and increases monotonically with increasing ***B*** without any signs of saturation, showing a typical metal or semimetal nature of high carriers. On the other hand, the MR increases with decreasing temperature and reaches ~3% at 2 K without saturation. This non-saturating MR is somewhat anomalous and cannot be explained using semiclassical theory for a trivial band structure [29]. Therefore, such results are closely linked with the nontrivial or surface bands, directly related to the topological features. The transverse resistivity ($\rho_{xy}$) (see Figure S6(b)) as a function of the magnetic field increases with increasing magnetic field with a positive slope, indicating the *p*-type conductivity, which can endorse the majority of the hole carrier density. Therefore, we have further estimated the carrier density and carrier mobilities as a function of temperature from the Hall resistivity data using two model fitting [30] $\sigma_{xy} = \left[\frac{n_e \mu_e^2}{1+\mu_e^2 B^2} + \frac{n_h \mu_h^2}{1+\mu_h^2 B^2}\right] eB$ (see Figure S6(c)). Where $n_e$ and $n_h$ show the carrier density of electrons and holes, while $\mu_e$ and $\mu_h$ indicate the carrier mobility of electrons and holes, respectively. The results indicated that the carrier concentration and mobility of both electrons and holes decreased with increasing temperature, having the hole concentration larger than the electron concentration at all tested temperatures, which reveals a typical strong metallic nature with the majority of the hole carrier density (see Figure S6(d)). The ARPES measurements further depict the Fermi level lying inside the valence band (see Figure 1). These results revealed that the carrier density such kinds of nodal-line semimetals is comparable with Dirac/Weyl semimetals [30, 31]. Whereas the carrier mobility of electrons and holes shows a decreasing trend with temperature for $Ta_3GeTe_6$ (see Figure S6(e)).

With the existence of the anisotropic surface features, we have further measured the nonreciprocal magnetoresistance on the fabricated device (see Figure 2(a)) using a four-probe configuration device with applied $I_{ac}$ = ± 1 mA (see Figure 2(e)). Figure 1(e) shows the nonreciprocal MR in response to both forward and backward charge flows, $I_{ac}$ = ± 1 mA, in a magnetic field sweep perpendicular to the *c*-axis. We can clearly see that with increasing ***B*** can detect a substantial resistance variation for the forward and reverse current flow. A higher resistance can be observed for the positive ***B*** with applied forward current flow, while the resistance becomes higher for negative ***B*** in the reverse current flow. The reverse MR effect can represent the directional charge transport in this thin flake $Ta_3GeTe_6$ sample. At the same time, the observed nonreciprocal MR, defined as ΔMR = MR(+*I*)–MR(−*I*), displays a linear increase under the magnitude of an applied magnetic field (see Figure S5(f)). The observed asymmetry of the MR is attributed to the nonreciprocal MR, which is further implicit in the current-induced effective Rashba-type charge-induced field normal to a current flow in the polar kind of system.

To interpret the experimentally observed current-direction-dependent magnetoresistance, we model the nonreciprocal transport within the framework of Onsager's reciprocal relations (Toy model). The electrical conductivity tensor for a system with diffusive charge carriers in an external magnetic field $\boldsymbol{B}$ can be written [32] as:

$$\sigma_{ij}(\boldsymbol{B}, \mathrm{k}) = \sigma_{ji}(-\boldsymbol{B}, -\mathbf{k}), \tag{1}$$

where $\mathbf{k}$ is the average wave vector of the diffusing particles. Expanding Eq. (1) in terms of the symmetry-breaking polarity vector $\mathbf{p}$ and $\boldsymbol{B}$, we obtain:

$$\sigma_{ij}(\boldsymbol{B}, \mathbf{k}) = \sigma_{ij}^{(0)} + \sigma_{ij}^{(1)} k . (\mathbf{p} \times \boldsymbol{B}) + \sigma_{ij}^{(2)} \boldsymbol{B}^2 + \cdots \tag{2}$$

where $\sigma_{ij}^{(0)}$ is the conductivity without $\mathbf{p}$ and $\boldsymbol{B}$, while $\sigma_{ij}^{(1)}$ and $\sigma_{ij}^{(2)}$ are first- and second-order contributions in $\boldsymbol{B}$, respectively. The average wave vector $\mathbf{k}$ is driven by the applied electrical current $\boldsymbol{I}$. By retaining terms up to second order in $\boldsymbol{B}$, the expression can be recast in terms of resistance as:

$$R(I, B, p) = R_0[1 + \chi \mathbf{p} . (\boldsymbol{B} \times \boldsymbol{I}) + \alpha \boldsymbol{B}^2] \tag{3}$$

where $p = 0$ corresponds to inversion-symmetric systems, and $p \neq 0$ describes inversion symmetry breaking. Here, $\chi$ denotes the nonreciprocal magnetoresistance coefficient, $I$ is the current, $\alpha$ is the symmetric magnetoresistance coefficient, and $R_0$ is the zero-field longitudinal resistance. Microscopically, the coefficient $\chi$ is intrinsically tied to the surface-state parameters and scales as $\chi \propto P_s \cdot \alpha_R$, where $P_s$ is the magnitude of the surface spin polarization and $\alpha_R$ represents the surface Rashba spin-orbit coupling strength. A larger $\alpha_R$ enhances the current-induced effective magnetic field $\left(\overrightarrow{B_{eff}} \propto \vec{v} \times \vec{E}\right)$ stemming from the surface inversion symmetry breaking ($p \neq 0$), while a higher $P_s$ maximizes the asymmetry in spin-dependent scattering channels when an external magnetic field is applied. Equation (3) shows that magnetoresistance depends on the relative orientation of the applied magnetic field and injected current. For inversion-symmetric systems $p = 0$, the MR response remains symmetric upon current reversal. In contrast, when inversion symmetry is broken ($p \neq 0$), the MR exhibits an asymmetric, nonreciprocal behavior. Figure 2(f) illustrates the results of this toy model, which aligns well with our experimental observations. Thus, both the DFT and phenomenological Toy model consistently validate that the unidirectional transport in $Ta_3GeTe_6$ originates from the synergy between strong SOC and surface inversion-symmetry breaking.

With such anisotropic charge flow in this vdW material, we further performed details magnetotransport measurements along with second harmonic experiments using DC ($I_{dc}$) and AC currents ($I_{ac}$). Figure 3(a) presents the typical schematic design of the electrodes on the sample with applied current and voltage direction for nonreciprocal charge transport measurement and simple transport experiments. The current can be applied to the *x*-axis, and a longitudinal voltage will be measured while applying a magnetic field. The Angle-dependent magnetoresistance ($\Delta R_{xx}$) of the device was measured at different temperatures using $\boldsymbol{B}$ = 9 T and $I_{dc}$ = 3 mA. The $\Delta R_{xx}$ can be obtained from the relation ($\Delta R_{xx} = R_{xx}(\theta) - R_{xx}(\theta = 0))/R_{xx}(\theta = 0)$, while rotating a device in the *xy* plane (see Figure 3(b)). The results indicated that the MR shows asymmetric behavior of directional dependent transport with the two-fold anisotropy of the device. The resistivity of the sample varies sinusoidally with respect to the angle, with the highest amplitude being observed at 90$^{\circ}$ while the lowest is at 0$^{\circ}$ and 180$^{\circ}$. The variation of resistivity with angle follows $R = \Delta R\sin\theta$, where $\Delta R$ is the amplitude, and $\theta$ is the rotation angle. With increasing temperature, the amplitude of the $\Delta R_{xx}$ decreases; however, the trend of the anisotropy exists up to 50 K. At the same time, we have measured the $\Delta R_{xx}$ as a function of theta at different $\boldsymbol{B}$ values from a positive $\boldsymbol{B}$ to a negative $\boldsymbol{B}$ (see Figure 3(c)). Interestingly, $\Delta R_{xx}$ flips when the magnetic field direction is changed, which corroborates the unidirectional magnetoresistance [22, 26, 27, 33, 34]. With decreasing $\boldsymbol{B}$, the amplitude of $\Delta R_{xx}$ is reduced for both positive and negative $\boldsymbol{B}$. Additionally, we have performed the measurements at varying currents, as shown in Figure 3(d). With increasing applied current, the $\Delta R_{xx}$ is monotonically increasing. These results also indicate that the asymmetric trend in response to the current with increasing differential resistance.

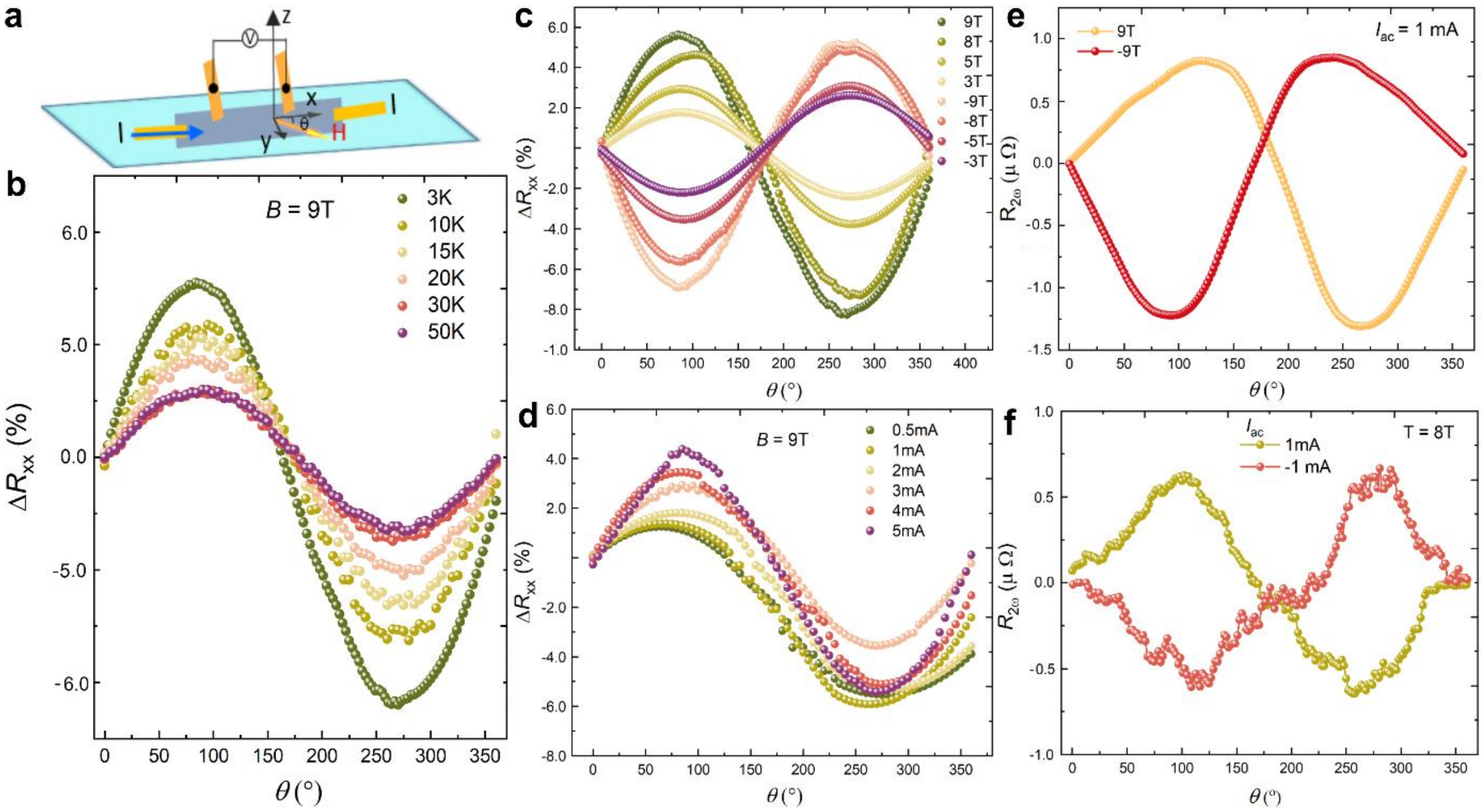

**Figure 3. Nonreciprocal charge transport in $Ta_3GeTe_6$: (**a) Schematic of the measurement configuration. (b) $\Delta R_{xx}$ having ***B*** rotating in *xy* plane measured at different temperatures (3 K to 50 K) and a fixed ***B*** = 9T, (c) $\Delta R_{xx}$ measured at different ***B*** and *T* = 5 K, the measurements (b & c) were performed at $I_{dc}$ = 3 mA. (d) $\Delta R_{xx}$ at increasing current ($I_{dc}$) and ***B*** = 9T and temperature of 5 K. Unidirectional magnetoresistance of thin flake device of $Ta_3GeTe_6$ (e) Angle-dependent $R_{2\omega}$ measured with a rotation of ***B*** = ± 9 T in the *yx* plane under applied $I_{ac}$ = 1 mA and *T* = 2 K. (f) Angular dependence of $R_{2\omega}$ under the opposite $I_{ac}$ (±1 mA) at ***B*** = 8T and *T* = 2K.

With these careful investigations of the unidirectional MR effect, we have further examined the device under applied AC current ($I = I_0\sin\omega t$), having phase detection for $\Delta R$. We have measured the nonreciprocal resistance ($R_{2\omega}$), which was directly measured from the out-of-phase component, as $R_{2\omega} = V_{2\omega}/I_0 = 1/2\ \gamma R_0 B I_0 \sin\theta$ [35]. Figure 3(e) exhibits the angle-dependent $R_{2\omega}$ measured at ***B*** = ± 9 T and at T = 2 K for $I_{ac}$ = 1 mA. It is observed that the sinusoidal resistance curves at out-of-plane rotational ***B*** = +9 T, with the maximum amplitude appearing at 90 degrees (+y) on the positive side and 270 degrees (-y) on the negative side. However, when changing the sign of the field, i.e. ***B*** = -9 T, one can reverse the angle-dependent $R_{2\omega}$ curve, illustrating the nonreciprocal unidirectional behaviour of MR at the AC. At the same time, we have measured the $R_{2\omega}$ by changing the phase of the AC current, which can also flip the sign; nevertheless, the amplitude is quite similar with a slight decrease due to the applied field strength (see Figure 3(f)). The $R_{2\omega}$ exhibits the angular dependence of the direction of the current with respect to the crystallographic axes of the sample. These results follow the relation of $R_{2\omega}(I, B) = - R_{2\omega}(-I, B) = -R_{2\omega} (I, -B) = R_{2\omega} (-I, -B)$, which can demonstrate the unidirectional magnetoresistance in thin $Ta_3GeTe_6$ that is odd concerning the polarity of either $I$ or $\boldsymbol{B}$ [22]. Our results clearly demonstrate that the magneto-transport study of $Ta_3GeTe_6$ shows the nonreciprocal charge transport characteristics, which are the signatures of the spin-orbit coupling and spin-polarized Rashba-type effect in such thin flake anisotropic systems. To further demonstrate the structural anisotropy, we have used the polarized Raman spectroscopy measurements (see Figure S7). The polarized Raman spectra clearly demonstrate the two-fold and four-fold phonon symmetries, reflecting the structural anisotropy of this crystal. These specific anisotropic phonon modes responsible for this directional electron scattering link the anisotropic structure with directional transport properties.

To validate and conclude the analysis, we have used the density functional theory (DFT) calculations, which reveal that $Ta_3GeTe_6$ crystallizes in a layered orthorhombic structure with space group $P_{\mathrm{nma}}$, composed of Ta–Ge–Ta zigzag chains sandwiched between the Te layers (Figure 4(a)). The bulk and (001) surface projected Brillouin zones are shown in Figure 4(b). The bulk band structure illustrated in Figure 4(c) exhibits Dirac-like dispersions near the Fermi level at the border of the Brillouin zone, confirming the semimetallic nature, revealing multiple symmetry-protected band crossings, which endorse our ARPES results. An hourglass Dirac dispersion is observed along the XS direction close to the Fermi level. Figures 4(d–f) displays the complex Fermi surfaces, further highlighting the anisotropic electronic topology, while surface-projected spectral functions shown in Figure 4(g) demonstrate topologically nontrivial states localized at the (001) surface. The spin-resolved band structure calculations in Figure 4(h) clearly reveal spin splitting and Rashba-type features indicating spin–momentum locking, directly resulting from inversion-symmetry breaking at the surface and strong SOC from Ta 5*d* orbitals. Such a Rashba-like band was also observed in the ARPES data in Figure 1(g).

These results establish a link between the electronic structure and the transport properties: the bulk of $Ta_3GeTe_6$ retains inversion symmetry, but the inversion symmetry breaking at the surface induces Rashba-type spin-polarized states, a clear signature enabling a nonreciprocal transport mechanism in the structure. Microscopically, as schematically illustrated in Figure 4(i), surface termination breaks $\mathcal{P}$-symmetry to produce an hourglass semimetal with spin-momentum locked helical Fermi surfaces (middle panel). When a longitudinal current *I* and an in-plane magnetic field $\mathrm{B} \perp \mathrm{I}$ are applied (right panel), the Zeeman field shifts these momentum-locked sub-bands, breaking the $+\mathrm{k}/-\mathrm{k}$ carrier degeneracy $\left(\varepsilon(\mathrm{k},\mathrm{B}) \neq \varepsilon(-\mathrm{k},\mathrm{B})\right)$. This momentum asymmetry induces non-equivalent carrier group velocities and scattering lifetimes, directly generating the second-harmonic nonreciprocal resistance $(R^{2\omega} \propto \mathrm{I} \cdot \mathrm{B})$. Collectively, these results establish $Ta_3GeTe_6$ as an hourglass semimetal with non-trivial topology and nonreciprocal transport behavior, making it a potential candidate for future quantum devices.

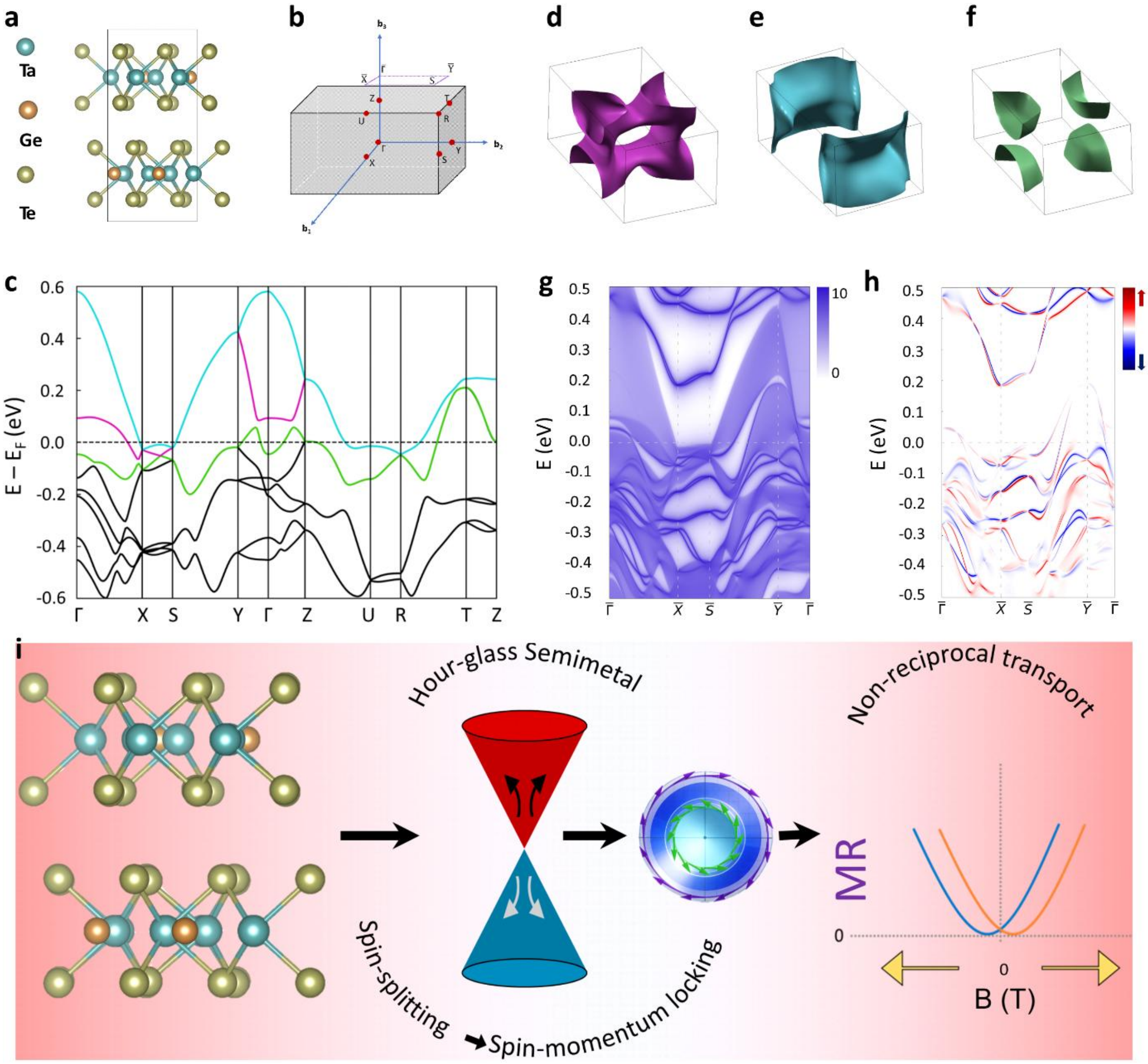

**Figure 4. Broken inversion symmetry and spin-polarized surface states calculations using DFT:** (a) Crystal structure of $Ta_3GeTe_6$. (b) Bulk Brillouin zone (BZ) with high-symmetry points and paths. (c) Electronic band structure showing multiple band-crossings near the Fermi level. (d–f) Fermi surface plots illustrating the complex and anisotropic electronic topology (g) Surface-projected electronic band structure, revealing nontrivial topological surface states. (h) Spin-resolved surface band dispersion illustrating spin–momentum locking features, where red and blue denote opposite spin orientations. The broken inversion symmetry at the surface produces spin-polarized surface states, enabling nonreciprocal charge transport. (i) Schematic overview of the inversion symmtry spin spliting and spin momentum locking produce nonreciprocal MR.

## 3. Conclusion

High-quality $Ta_3GeTe_6$ single crystals were successfully synthesized. After characterization, the structural and electronic anisotropic nature of these single crystals was unambiguously proved by using ARPES, magneto-transport, and Raman spectroscopy experiments. ARPES has unequivocally proved the non-trivial topological and Dirac-like hourglass semimetal these single crystals. Transport properties in combination with ARPES have also shown that $Ta_3GeTe_6$ possesses hole-type carriers. Additionally, magneto-transport studies on $Ta_3GeTe_6$ show non-saturating magneto-resistance, which changes with an applied magnetic field and depicts the phenomenon of unidirectional magnetoresistance along with structural anisotropy was demonstrated by Raman spectroscopy. The experimental observations were complemented with the first-principles DFT calculations, revealing that $Ta_3GeTe_6$ is an hourglass semimetal with spin-polarized surface states and broken inversion symmetry on the surface, providing the microscopic origin of nonreciprocal charge transport observed experimentally. Moreover, the toy model quantitatively captures the nonreciprocal resistance term, which is in excellent agreement with the measured direction-dependent magnetoresistance. Together, these theoretical and experimental findings establish $Ta_3GeTe_6$ as a platform for exploring nonreciprocal transport and anisotropic quantum phenomena in topological semimetals.

**Conflict of Interest**

The authors declare no competing financial interests.

**Data availability**

The raw data required to reproduce the above findings are available to download from https://doi.org/10.58132/6XKAS5.

**Acknowledgments**

The project was supported by Beijing Natural Science Foundation (No. 4232070), the National Natural Science Foundation of China (No. W2632012, 1261101385, T2394475 and 52261145694), and the International Mobility Project (No. B16001). I. A. and M. R. M. acknowledge the support from the National Science Centre, Poland (Grant No. 2022/46/E/ST3/00166). C. A. was supported by the "MagTop" project (FENG.02.01-IP.05-0028/23) carried out within the "International Research Agendas" programme of the Foundation for Polish Science co-financed by the European Union under the European Funds for Smart Economy 2021-2027 (FENG). We also acknowledged beamline 13U of the National Synchrotron Radiation Laboratory (NSRL) for the ARPES experiments.

# Supporting Information

## S1. Experiments and Methods

***Bulk crystal synthesis:***

Bulk crystal synthesis: $Ta_3GeTe_6$ single crystals were synthesized by using CVT. The precursor materials of Ta, Ge and Te, with a 3: 2: 6 stoichiometric ratio (total mass of 2 mg), were mixed and ground together to form a homogeneous mixture, with a slight excess of Te (less than 5 wt.%) added to compensate for its volatility during the high-temperature growth process. Then the mixture was transferred to a silica ampoule with an inner diameter of 10 mm, a wall thickness of 2 mm, and a length of 20 cm (internal volume $\approx 15.71$ cm$^3$). Resublimed iodine ($I_2$) with a mass of 35 mg $(\sim 2.23\ \mathrm{mg/cm^3})$ was added to act as the transport agent. The ampoule was evacuated and sealed under a vacuum pressure of $10^{-4}$ Torr. Thereafter, the ampoule was loaded coaxially into a two-zone furnace, and the temperature of the furnace was raised with a gradient of $1\ ^\circ\mathrm{C/minute}$ until both its zones reached 750 °C. It was kept at this temperature for 6 days and was subsequently reduced to get a high-temperature zone at 600 °C, while the low-temperature zone was maintained at 550 °C. The furnace was allowed to operate at this temperature for the next 7 days before naturally cooling to room temperature. This experiment yielded high-quality single crystals of $Ta_3GeTe_6$, which were utilized for further investigation thereafter.

***ARPES Measurement:***

Angle-resolved photoemission spectroscopy (ARPES) measurements on $Ta_3GeTe_2$ were performed at the BL13U beamline of the National Synchrotron Radiation Laboratory (NSRL) in Hefei. Using a Scienta DA30L electron spectrometer and horizontally polarized light with photon energies from 20 to 33 eV, the electronic structure was mapped. The sample was cleaved in situ at 12 K and measured at 8 K in an ultrahigh vacuum better than $1.6 \times 10^{-11}$ mbar. The energy and angular resolutions were 15 meV and 0.2°, respectively.

***Device fabrication and transport measurements:***

The High-quality $Ta_3GeTe_6$ flakes were mechanically exfoliated from bulk crystals using white Scotch Magic tape. The exfoliated layers were first transferred onto a flexible polydimethylsiloxane (PDMS) substrate, and the uniform contrast and suitable thickness were identified under an optical microscope. The selected layers were sequentially transferred onto a pre-patterned Hall bar with a width of 5 μm (prepared by standard photolithography technique and Ar ion milling with a Ti(10 nm)/Au(50 nm) electrodes). The device was wire-bonded to a rotatable sample holder using aluminum wires and loaded into a Physical Property Measurement System (PPMS, Quantum Design) for transport measurements. A sinusoidal AC current at 13 Hz was supplied by a Keithley 6221 current source. The resulting first-harmonic and second-harmonic voltage signals were measured using two Stanford Research SR830 lock-in amplifiers, configured to detect the in-phase (0°) and out-of-phase (90°) components, respectively. By shifting the phase reference of the lock-in amplifier by 180°, we successfully isolated the phase-resolved directional transport characteristics corresponding to the effective forward ($I_{ac} > 0$) and backward ($I_{ac} < 0$) current directions, as shown in Fig. 2(e).

***Raman Measurements:***

Polarization-resolved Raman scattering measurements were conducted at ambient temperature using a 515 nm diode laser with a power of 2 mW. Reflected laser light and Rayleigh scattering were suppressed using a 515 nm Bragg-grating notch filter. For the angle-dependent measurements, a λ/2 plate was placed after the objective to simultaneously rotate the polarization of both the incident and scattered light with respect to the crystal axes, while maintaining propagation along the sample's z-axis. Raman spectra were collected in both co-polarized (XX) and cross-polarized (XY) configurations.

***Density functional theory calculations:***

Density functional theory (DFT) calculations were carried out using the VASP package [1]. The core and valence electrons are considered within the projector augmented wave method by taking a cutoff of 600 eV for the plane wave basis [2, 3]. For structural optimization, both the volume and atomic position relaxations are performed until the force and energy differences were less than 0.001 eV/Å and $10^{-8}$ eV, respectively. An appropriate $k$-mesh $8 \times 6 \times 4$ was chosen for each structure depending on the lattice parameters. All the electronic band structure calculations were performed by considering the spin-orbit coupling (SOC) effects. The calculations for the surface states were performed using the semi-infinite Green's function approach integrated in the WannierTools [4, 5].

## S2. Supporting results

### S2.1. Characterizations

Figure S1 shows the characterization results of $Ta_3GeTe_6$ single crystals. Figure S1(a) illustrates the single-crystal X-ray diffraction (XRD) of the prepared single crystals. XRD exhibits sharp diffraction peaks, indicating the high quality of the crystalline nature of the grown single crystals, with no impurity peaks, suggesting that the crystals were grown in the pure phase. The crystals have grown along the (0X0) direction, with a preferred growth orientation along the (080) direction. Figure S1(b) shows the integrated core-level photoemission spectrum of the cleaved surface of the $Ta_3GeTe_6$ single crystal. It is seen that the spectrum displays Ta 4f, Ge 3d and Te 4d spin-orbit splitting doublets, which can confirm the chemical structure of the $Ta_3GeTe_6$ crystal. The Ta 4f states lie within 22 and 26 eV, originating from the Ta surface's chemical shift, and Ge 3d states occupy around 30 eV. The peak within 39 to 42 eV is allocated to the Te 4d orbitals, attributed to the Te surface.  The energy-dispersive X-ray spectroscopy (EDS) and elemental mapping were performed on the single crystal (see SEM image in Figure S1(a)) with corresponding EDS spectra (see Figure S1(b)) and their mapping elucidating that all three elements (Ta, Ge, and Te) are uniformly distributed throughout the crystals without any aggregation (see Figure S1(c)). The EDS spectra taken from the $Ta_3GeTe_6$ surface in Figure S1(b) depict the stoichiometric ratio of the constituent elements i.e., Ta, Ge, and Te, respectively, which are nearly 3:1:6.

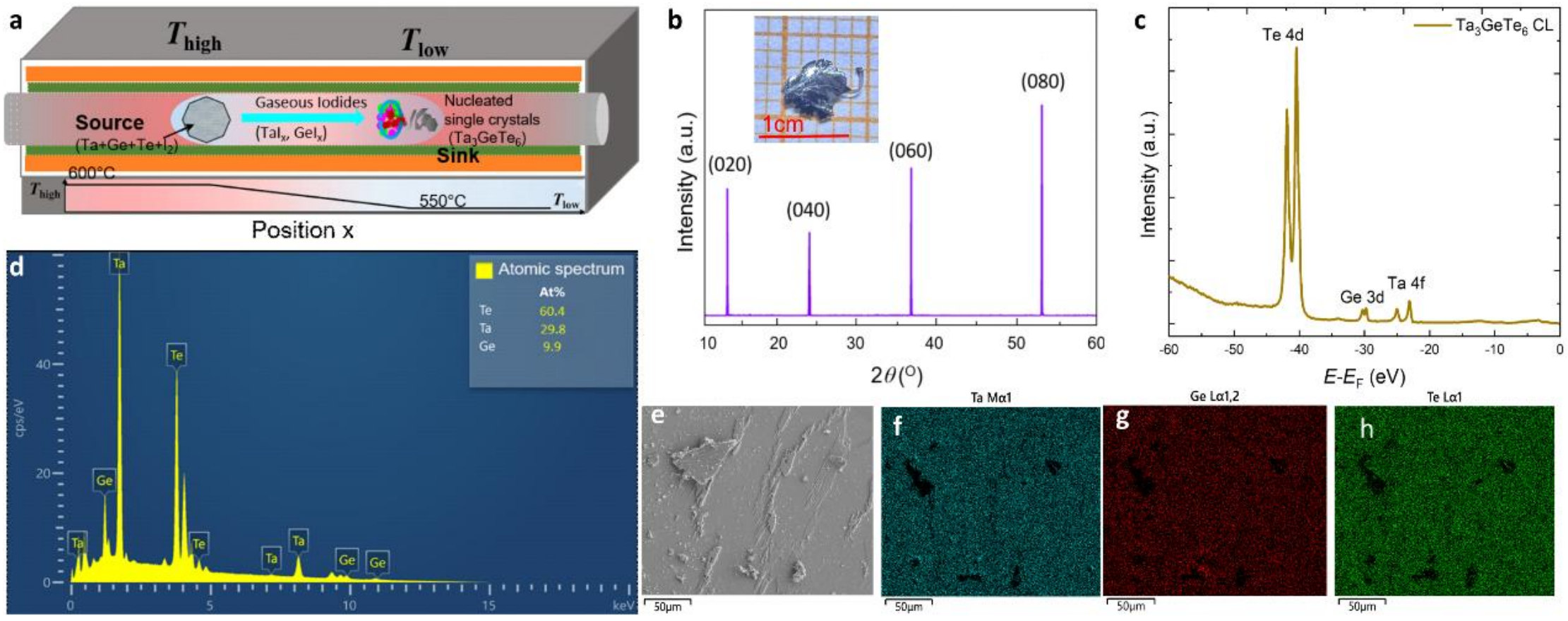

**Figure S1.** Synthesis and characterization of $Ta_3GeTe_6$ bulk crystal. (a) schematic of the CVT furnace used to grow the $Ta_3GeTe_6$ bulk crystal growth via iodine vapor transport. (b) Single-crystal X-ray diffraction (XRD) pattern of the as-grown $Ta_3GeTe_6$ crystal, showing sharp peaks corresponding to the (0*l*0) family of planes. The inset shows an optical photograph of a typical as-grown single crystal placed on a millimeter-grid background (scale bar is 1 cm), showcasing its characteristic morphology and metallic luster. (c) Core level spectra of the $Ta_3GeTe_6$, showing the Ta, Ge and Te peaks. (d) EDS spectrum of $Ta_3GeTe_6$, with the inset table showing the obtained composition from the spectrum data of the region from the SEM image (e) and their corresponding Mapping images show Ta, Ge and Te, respectively.

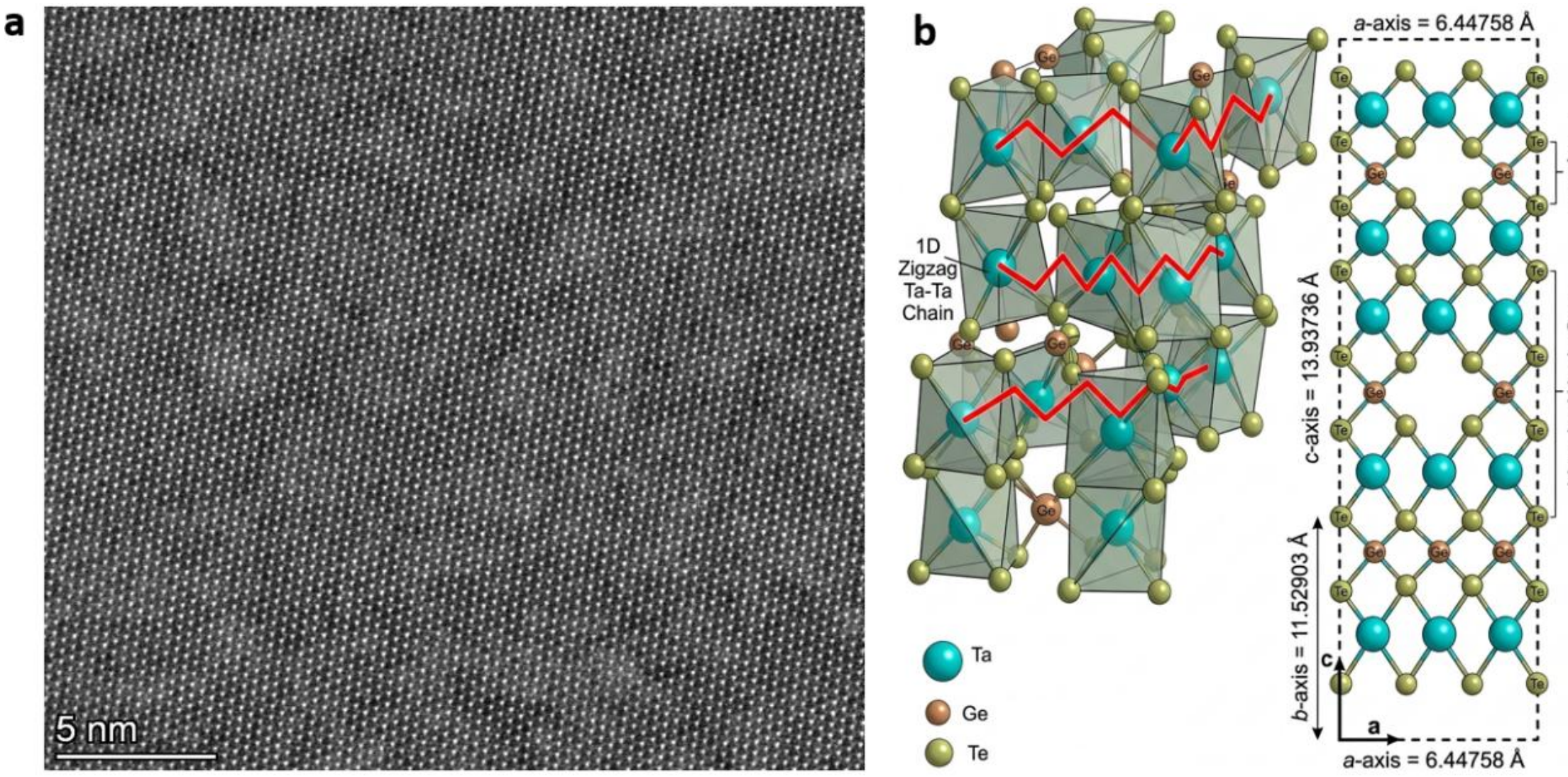

**Figure S2:** (a) Microstructure of the $Ta_3GeTe_6$ using STEM measurement and (b) crystal structure representation using base vectors and vdW bonding, making the zigzag chain.

### *S2.2. ARPES Measurement*

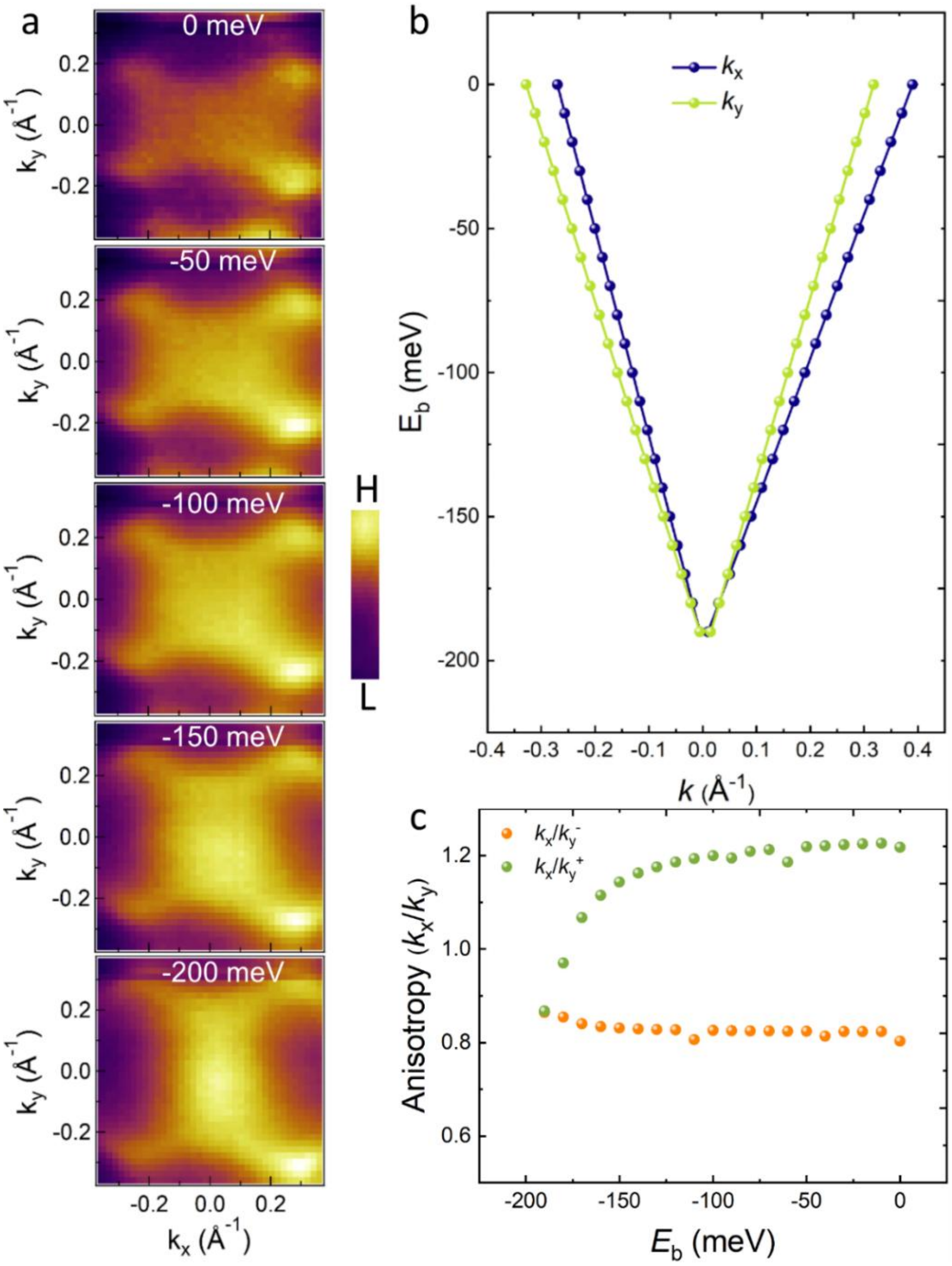

**Figure S3:** (a) CEC plots at different binding energies, (b) $k_x$ and $k_y$ representation as a function of different binding energies and (c) shows the calculated anisotropy representation of the FS at different binding energies. unequivocally

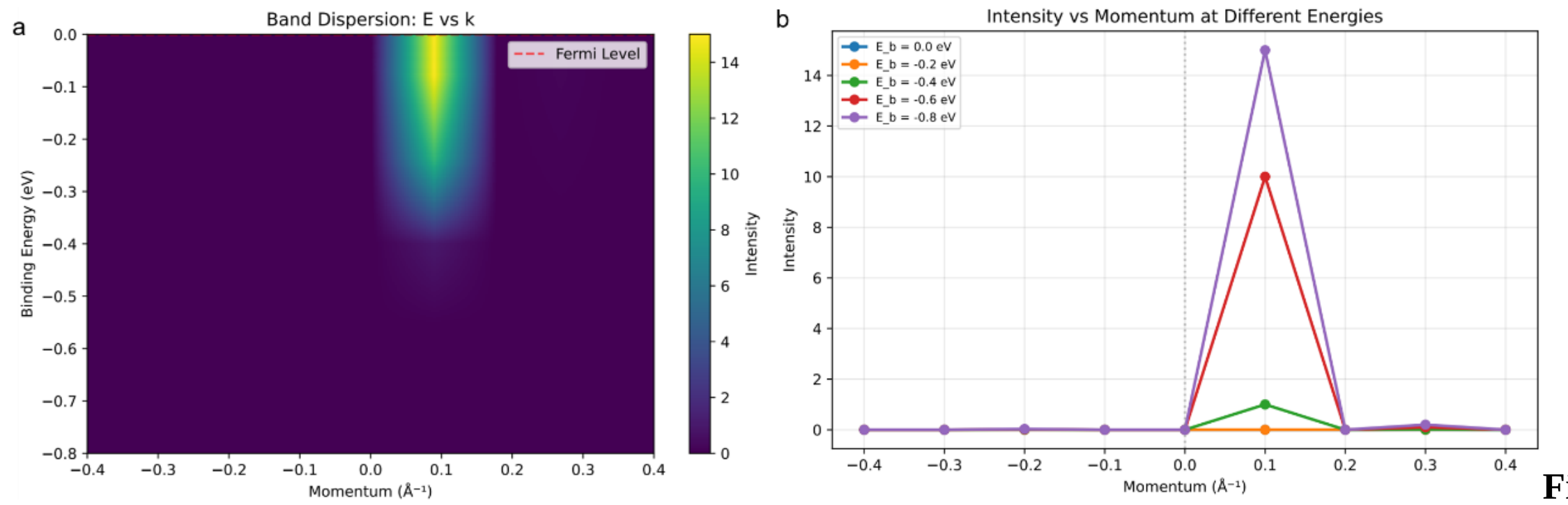

**Figure S4: Nodal line analysis:** (a) Figure S5: Single line shows the Nodal line dispersion along *k*-space and (b) variation of intensity as a function of binding energies, respectively.

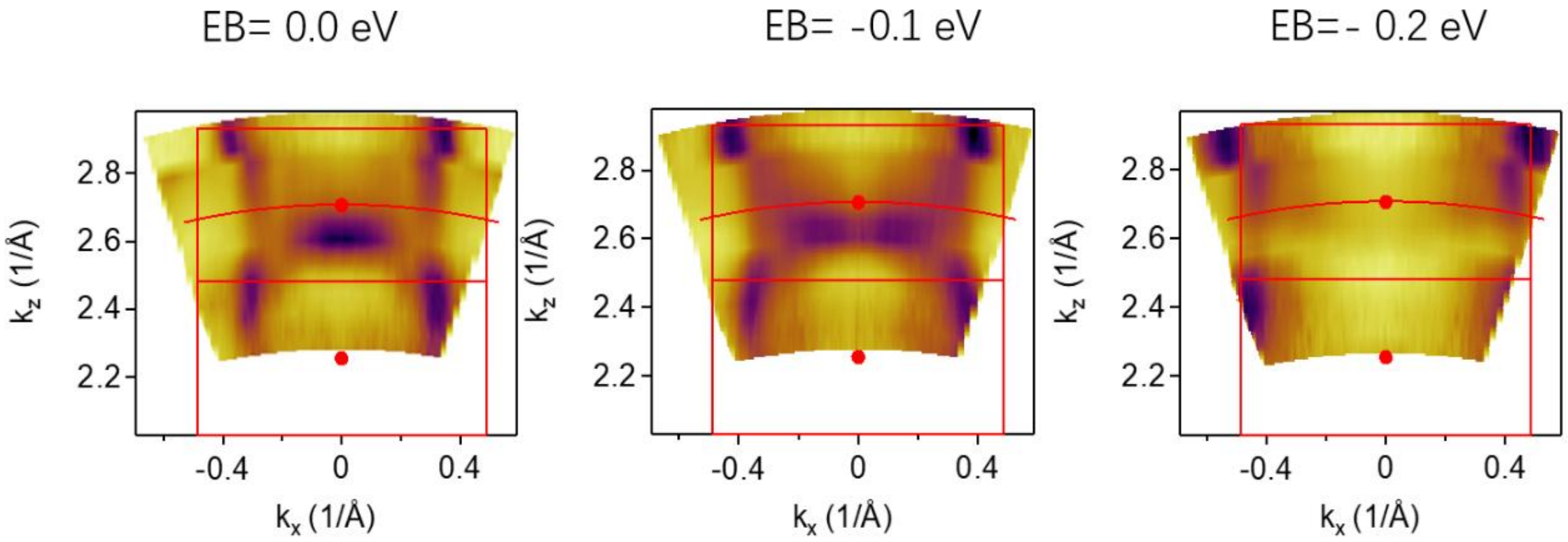

**Figure S5:** Photon energy dependence contour plots along $k_x$ and $k_z$ of $Ta_3GeTe_6$. The measurements was taken from 20 eV to 33 eV.

## *S2.3. Transport Results:*

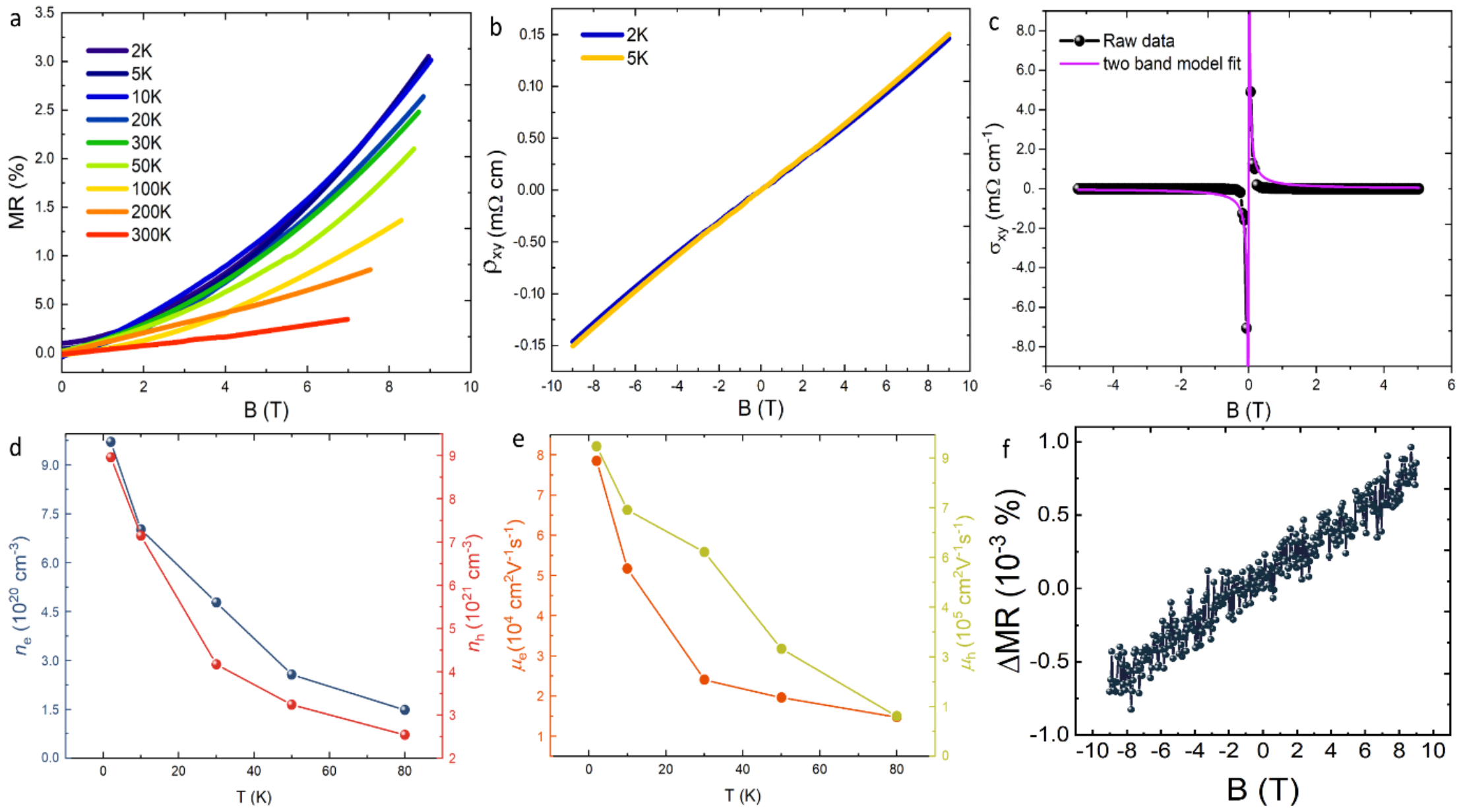

**Figure S6: Transport measurements and analysis:** (a) MR at different temperatures, (b) transverse resistivity ($\rho_{xy}$) at 2 K and 5 K, and (c) two-band model fitting for 2 K. (d & e) carrier concentration and mobility of electrons and holes as a function of temperature, respectively. (f) The magnitude of nonreciprocal MR linearly increases with increasing applied *B*.

### *S2.4. Polarized Raman spectroscopy*

The phonon dispersion of the bulk $Ta_3GeTe_6$ sample was studied using angle-resolved polarized Raman scattering (ARPRS) to demonstrate the anisotropic features of phonon dispersion related to the crystal orientation. It is indicated that the $Ta_3GeTe_6$ shows a very rich Raman spectrum of 21 modes from the 70 to 180 cm$^{-1}$ range. The ARPRS was measured under two different polarization configurations to explore the phonon anisotropy in our grown crystal. Figure S6(b) and (c) represent the contour plots of ARPRS for co-(XX)-and cross-(XY)-polarization configurations. It is affirmed from both polarization configurations that the RS shows different intensities at varying angles. At XX configuration, each mode of RS below 130 cm$^{-1}$ shows two-fold symmetry except the modes at 87 and 103 cm$^{-1}$, while, for XY configuration the RS illustrates four-fold symmetry for all modes apart from the mode at 92 cm$^{-1}$. At the same time, the Raman mode 180 cm$^{-1}$ is not observed in the XY configuration, indicating an out-of-plane symmetry of this mode. Furthermore, we have analyzed the Raman modes above 130 cm$^{-1}$ due to unclear intensities of these modes for both polarization by using polar plots of the integrated intensities, as shown in Figures S6(d-i). Figure S6(d-i) represents the polar plots of Raman intensity and fitted curves (grey dashed line) of the observed six Raman-active modes above 130 cm$^{-1}$ for both co-and cross-polarization, respectively. The dashed curves in Figures S6(d-i) depict the fits of the mode intensities as a function of light polarization, $I(\theta)$, by the relation [6].

$$I(\theta) = (|A|\sin^2(\theta - \phi) + |C|\cos(\alpha)\cos^2(\theta - \phi))^2 + |C|^2\sin^2(\alpha)\cos^4(\theta - \phi) \quad (1)$$

where |A| and |C| indicate the amplitudes of the phonon modes, $\phi$ denotes the phase of polarization dependence, and $\alpha$ is the phase difference. The results indicate that the four modes under XX-polarization show 2-fold symmetry (see Figures 4(d-g)), and two modes (163 and 166 cm$^{-1}$, see Figure S6(h & i)) illustrate 4-fold symmetry. Whereas, for XY-polarization configuration, all the Raman modes above 130 cm$^{-1}$ show 4-fold symmetry except the mode at 149 cm$^{-1}$ (see Figure S6(f)) doesn't show any polarization dependence. The Raman results demonstrate that the phonons are anisotropic for both XX and XY polarization configurations, which is associated with the crystal symmetry of $Ta_3GeTe_2$. Our results demonstrate that not only the electronic but also the phononic properties of this van der Wall system are anisotropic, which can be associated with our crystal structure, which can show the unidirectional magnetoresistance.

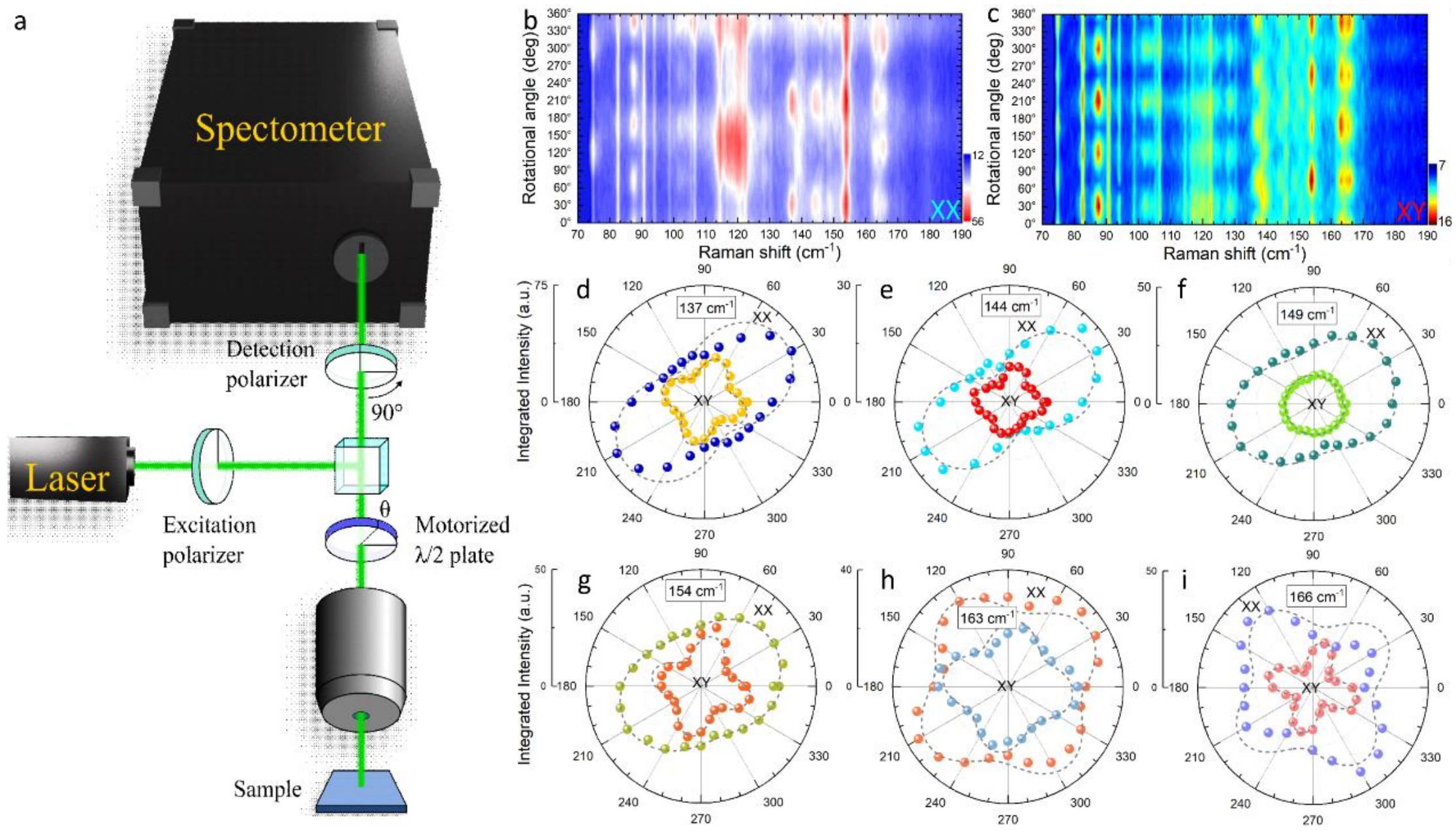

**Figure S7: RS of the Bulk $Ta_3GeTe_6$:** (a) Schematic illustration of the Raman measurement (b, c) contour plots of the ARPRS and (d-i) polarization dependence of the integrated intensity of the Raman modes (137, 144, 149, 154, 163 and 166 $cm^{-1}$) for co-and cross-polarization configurations, respectively. The grey dashed line shows the fitting.